\documentclass[12pt,letterpaper]{article}
\pdfoutput=1
\usepackage{jheppub}
\usepackage{amsfonts, amsthm}
\usepackage[english]{babel}
\usepackage[utf8]{inputenc}
\usepackage{slashed}
\usepackage{mathrsfs}
\usepackage{amssymb}
\usepackage{color}
\hypersetup{unicode}

\newcommand{\eq}{\begin{equation}}
\newcommand{\feq}{\end{equation}}
\newcommand{\eqn}{\begin{eqnarray}}
\newcommand{\feqn}{\end{eqnarray}}

\newcommand{\ma}[1]{\mbox{$\mathcal{#1}$}}

\newcommand{\masf}[1]{\mbox{$\mathsf{#1}$}}
\newcommand{\maf}[1]{\mbox{$\mathfrak{#1}$}}

\title{Nonlinear symmetries of black hole entropy in gauged supergravity}

\author{Dietmar Klemm$^{1}$,}
\author{Alessio Marrani$^{2,3}$,}
\author{Nicol\`o Petri$^{1}$}
\author{and Marco Rabbiosi$^{1}$}

\affiliation{$^1$ Dipartimento di Fisica, Universit\`a di Milano, and \\
INFN, Sezione di Milano,\\
Via Celoria 16, I-20133 Milano, Italy}

\affiliation{$^2$ Museo Storico della Fisica e Centro Studi e Ricerche `Enrico Fermi',\\ Via Panisperna 89A, I-00184 Roma, Italy}

\affiliation{$^3$ Dipartimento di Fisica e Astronomia `Galileo Galilei', Universit\`a di Padova, and \\
INFN, Sezione di Padova,
Via Marzolo 8, I-35131 Padova, Italy}

\emailAdd{dietmar.klemm@mi.infn.it}
\emailAdd{Alessio.Marrani@pd.infn.it}
\emailAdd{nicolo.petri@mi.infn.it}
\emailAdd{marco.rabbiosi@mi.infn.it}

\preprint{DFPD/2016/TH-12\\ \hspace*{\fill} IFUM-1053-FT}

\abstract{Freudenthal duality in $N=2$, $D=4$ ungauged supergravity is generated by an
anti-involutive operator that acts on the electromagnetic fluxes, and results to be a symmetry of the
Bekenstein-Hawking entropy. We show that, with a suitable extension, this duality can be generalized to
the abelian gauged case as well, even in presence of hypermultiplets.
By defining Freudenthal duality along the scalar flow, one can prove that two configurations of charges and gaugings linked by the Freudenthal operator share the same set of values of the scalar fields at the black hole horizon. Consequently, Freudenthal duality is promoted to a nonlinear symmetry of the black hole entropy. We explicitly show this invariance for the model with prepotential $F=-i X^0 X^1$ and Fayet-Iliopoulos 
gauging.}


\begin{document}
\maketitle
\flushbottom

\section{Introduction}

A consistent understanding of the microscopic origin of black hole
entropy, and its relation to the macroscopic interpretation based on the
Bekenstein-Hawking entropy-area formula, should be a key feature
of any conceivable theory of quantum gravity. In this respect, string theory
is successful, even if its accomplishments are currently limited mainly to extremal
black holes that are asymptotically flat. In this framework,
the black hole solutions, pertaining to the field theory limit in which
supergravity arises, are described by configurations of wrapped D-brane
states. The microscopic origin of the entropy then arises from state
counting in a weakly coupled D-brane setup.

An important feature of supergravity black holes is the so-called attractor
mechanism \cite{Ferrara:1995ih,Strominger:1996kf,Ferrara:1996dd,Ferrara:1996um,Ferrara:1997tw}, describing the stabilization of the scalar fields near the event horizon only in terms of
the conserved charges of the system, regardless of the initial conditions
(asymptotic moduli) specified for their flow dynamics. This implies that
the entropy does not depend on the asymptotic values of the scalar fields.
After its discovery in asymptotically flat black holes in ungauged $N=2$, $D=4$ supergravity,
the attractor mechanism has then been extended to the presence of abelian
$\text{U}(1)$ Fayet-Iliopoulos (FI) gauging \cite{Bellucci:2008cb,Cacciatori:2009iz, Dall'Agata:2010gj}
and to nontrivial hypermultiplets \cite{Chimento:2015rra}, where an abelian subgroup of the
isometries of the quaternionic manifold was gauged.

Recent years have been characterized by an intense study of extremal black
holes in gauged supergravity. This led e.g.~to a symplectically covariant formulation of the equations
satisfied by the solutions, of their attractor mechanism and scalar flow dynamics, as well as to the
inclusion of coupling to hypermultiplets, cf.~e.g.~\cite{Chimento:2015rra,Hristov:2010eu,Hristov:2010ri,
Klemm:2012yg,Toldo:2012ec,Hristov:2012nu,Klemm:2012vm,Gnecchi:2012kb,Chow:2013gba,
Gnecchi:2013mja,Klemm:2015xda} for an (incomplete) list of references. In contrast with the case of
ungauged theories in which hypermultiplets can always be consistently decoupled, in gauged supergravity
hyperscalars may be charged and they actively participate to the solution.
After \cite{Hristov:2010eu}, in which new solutions in gauged supergravities
with nontrivial hypermultiplets were constructed by embedding known
solutions of ungauged theories, further advances were made in \cite{Halmagyi:2013sla}, where the gauged 
supergravity analogue of the BPS attractor equations for theories coupled to hypermultiplets are derived
and black holes with running hyperscalars are obtained numerically.
In \cite{Chimento:2015rra} the generalization of the effective black hole potential
formalism \cite{Bellucci:2008cb} to abelian gaugings of the quaternionic hyperscalar
manifold was given and in \cite{Klemm:2016wng} a symplectically covariant formulation of the
attractor mechanism and scalar flow dynamics in such a framework was achieved.

It is here worth remarking that hypermultiplets are essential in the formulation of realistic models given
by the low energy limit of string/M-theory flux compactifications, which in turn are one of the most
important motivations for the analysis of black hole solutions in gauged
supergravity theories. Flux compactifications are indeed an effectively
successful way to deal with the crucial moduli stabilization problem in
string theory. The fluxes give rise to a nontrivial gauge potential
in the effective theory, as well as to deformations determining gauged
supergravity models in the low energy limit \cite{Polchinski:1995sm,Michelson:1996pn,DallAgata:2001brr}.
Thus, the study of the attractor mechanism within this scenario
is of utmost importance \cite{Hsu:2006vw,Danielsson:2006jg}, because the presence of a charged
black hole may drive the value of the moduli fields to an attractor horizon
value which differs from the one obtained by the potential generated by flux
compactification in the asymptotic region.

Recently, a novel symmetry was discovered for black holes
in four-dimensional Einstein-Maxwell systems coupled to nonlinear sigma
models of scalar fields (which can be regarded as the purely bosonic sector
of an ungauged $D=4$ supergravity theory), namely the Freudenthal
duality. It can be defined as an anti-involutive, nonlinear map
acting on symplectic spaces, in particular on the representation space in
which the electromagnetic charges of the black holes sit.
After its introduction in \cite{Borsten:2009zy} in the context of the
so-called $U$-duality Lie groups of type E$_7$ \cite{Brown:1969} in extended supergravity theories, 
interesting relations between Freudenthal duality, the Hessian matrix of the black hole entropy and the
rigid special (pseudo-)K\"ahler metric of the prehomogeneous vector spaces
associated to the $U$-orbits, were discovered and studied in \cite{Ferrara:2013zga,Marrani:2015wra}.
In \cite{Ferrara:2011gv} Freudenthal duality
was proved to be a symmetry not only of the classical Bekenstein-Hawking
entropy, but also of the critical points of the black hole potential.
Moreover, it was consistently extended to any generalized special
geometry, thus encompassing all $N>2$ (extended) supergravities, as well as
$N=2$ generic special geometry, not necessarily having a coset space
structure.

Interestingly, Freudenthal duality made its appearance also in a number of
other contexts, such as gauge theories with symplectic scalar manifolds \cite{Marrani:2012uu}
and multi-centered black holes \cite{Fernandez-Melgarejo:2013ksa}. Moreover,
Lagrangian densities exhibiting Freudenthal duality as an on-shell symmetry
were constructed in \cite{Borsten:2012pd} (in the context of black hole solutions in $N=2$, $D=4$
supergravity, see also \cite{Galli:2012ji}).

All the above formulations and results on Freudenthal duality were confined
to ungauged theories. The present investigation is devoted to the
consistent formulation of Freudenthal duality in the context of abelian gaugings
of $N=2$, $D=4$ supergravity. This is done both for $\text{U}(1)$ FI gauging and for theories
coupled to hypermultiplets. As will be evident from the treatment
given below, an essential ingredient for such generalizations is the
effective black hole potential formalism introduced in \cite{Bellucci:2008cb,Chimento:2015rra}.

In particular, Freudenthal duality will be proved to be an
intrinsically nonlinear symmetry of the Bekenstein-Hawking extremal black
hole entropy. Besides generalizing the correponding result in ungauged
theories \cite{Borsten:2009zy, Ferrara:2011gv}, this enlarges the set of
invariance symmetries of the entropy function, thereby setting up the
challenging question of the realization/interpretation of such an
intrinsically non-linear symmetry in string/M-theory, also in the framework
in which a Dirac-Zwanziger-Schwinger quantization condition for dyonic
charges holds, and the symplectic representation space of electromagnetic fluxes
is turned into a charge lattice \cite{Borsten:2009zy}.

The remainder of this paper is organized as follows:
Section \ref{attr-gauged} contains a brief review of the attractor mechanism in
$N=2$, $D=4$ gauged supergravity coupled to vector- and hypermultiplets.
Section \ref{F-duality} is devoted to the introduction of Freudenthal
duality, starting with a summary of the ungauged case in \ref{sec:ungauged}. The extension to
$\text{U}(1)$ FI gauging is considered in section \ref{sec:FI}, and the further generalization to the coupling to
hypermultiplets is presented in \ref{hypers}. We conclude in sec.~\ref{conclusion} with some final remarks,
hinting to further future developments.

Throughout this paper, we use the conventions of \cite{Klemm:2016wng}.

\section{\label{attr-gauged}Attractors in gauged supergravity}

In ungauged supergravity, the attractor mechanism \cite{Ferrara:1995ih,Strominger:1996kf,Ferrara:1996dd,
Ferrara:1996um,Ferrara:1997tw} essentially states that, at the horizon of an
extremal black hole, the scalar fields $\phi$ of the theory are always attracted to the same values
$\phi_{\text h}$ (fixed by the black hole charges $\mathcal Q$), independently of their values
$\phi_\infty$ at infinity. When the so-called black hole potential has flat directions, it may happen that
some moduli are not stabilized, i.e., their values at the horizon are not fixed in terms of the black hole charges. Yet, the Bekenstein-Hawking entropy turns out to be independent of these unstabilized moduli. Notice that this does not hold anymore for nonextremal black holes, for which the horizon is not necessarily an attractor point. The $\phi_{\text h}$ are critical points of the black hole potential
$V_{\text{BH}}(\mathcal{Q},z^i)$, where in $N=2$, $D=4$ supergravity the $z^i$ denote only
the scalars in the vector multiplets, since hypermultiplets can be consistently decoupled.
The horizon values $z_{\text h}^i(\mathcal Q)$ are thus determined by the criticality conditions
\begin{equation}
\partial _i V_{\text{BH}}(\mathcal Q, z^i)|_{z_{\text h}^i(\mathcal Q)} = 0\,, \label{eq:critun}
\end{equation}
and the Bekenstein-Hawking entropy is given by
\begin{equation}
S_{\text{BH}} = \pi V_{\text{BH}}(\mathcal Q, z^i)|_{z_{\text h}^i(\mathcal Q)}\,. \label{SBH}
\end{equation}
In gauged supergravity, the scalar fields generically have a potential $V$, which contributes to
the $\phi_{\text h}(\mathcal Q)$ as well. Both for $\text{U}(1)$ Fayet-Iliopoulos gauging
\cite{Bellucci:2008cb} and for abelian gauged hypermultiplets \cite{Chimento:2015rra}, the black hole
potential in \eqref{eq:critun} has to be replaced by the effective potential
\begin{equation}
V_{\text{eff}} = \frac{\kappa - \sqrt{\kappa^2 - 4V_{\text{BH}} V}}{2V}\,, \label{eq:effective}
\end{equation}
where $\kappa=0,1,-1$ corresponds to flat, spherical and hyperbolic horizons respectively.
The limit for $V\to 0$ of $V_{\text{eff}}$ only exists for $\kappa=1$, in which case
$V_{\text{eff}}\to V_{\text{BH}}$, so one recovers correctly the black hole potential that governs the
attractor mechanism in ungauged supergravity. The fact that this limit does not exist for $\kappa=0,-1$
is not surprising since flat or hyperbolic horizon geometries are incompatible with vanishing scalar
potential. As before, the critical points of the effective potential determine the horizon values of the
moduli,
\begin{equation}
\partial_i V_{\text{eff}}(\mathcal Q, q^u, z^i)|_{z_{\text h}^i,\, q_{\text h}^u} = 0\,, \qquad
\partial_u V_{\text{eff}}(\mathcal Q, q^u, z^i)|_{z_{\text h}^i,\, q_{\text h}^u} = 0\,, \label{eq: crithyp}
\end{equation}
($q^u$ are the hyperscalars), and the entropy density reads
\begin{equation}
s_{\text{BH}} \equiv\frac{S_{\text{BH}}}{\text{vol}(\Sigma)} = \frac{V_{\text{eff}}(\mathcal Q, q^u,
z^i)|_{z_{\text h}^i,\,q_{\text h}^u}}4\,, \label{entr-density}
\end{equation}
where $\Sigma$ denotes the unit $\text{E}^2$, $\text{S}^2$ or $\text{H}^2$.

\section{\label{F-duality}Freudenthal duality}

In this section we shall briefly review the Freudenthal duality in ungauged
supergravity \cite{Borsten:2009zy,Ferrara:2011gv,Ferrara:2013zga}, and subsequently generalize
it to the gauged case.

\subsection{\label{sec:ungauged}Ungauged supergravity}

Following \cite{Ferrara:2011gv}, we introduce the scalar field dependent Freudenthal duality operator
$\mathfrak{F}_z$ by
\begin{equation}
\mathfrak{F}_z(\mathcal Q) \equiv \hat{\mathcal Q} = -\Omega\mathcal{M}\mathcal{Q}\,, \qquad
\mathfrak{F}_z(\mathcal V) \equiv \mathcal{V}\,, \label{eq:actF}
\end{equation}
where $\mathcal Q$ denotes the symplectic vector of charges, while the covariantly holomorphic
symplectic section $\mathcal V$ and the matrices $\mathcal M$, $\Omega$ were defined in
eqns.~(2.1), (2.7) and (2.9) of \cite{Klemm:2016wng} respectively. They satisfy the relations
\begin{equation}
{\mathcal M}^t = {\mathcal M}\,, \qquad {\mathcal M}\Omega{\mathcal M} = \Omega\,, \qquad
{\mathcal M}{\mathcal V} = i\Omega{\mathcal V}\,, \qquad {\mathcal M}D_i{\mathcal V} = -i\Omega 
D_i{\mathcal V}\,, \label{eq:Mprop}
\end{equation}
with $D_i$ the K\"ahler-covariant derivative. Moreover, the black hole potential can be written
in terms of $\mathcal Q$ and $\mathcal M$ as
\begin{equation}
V_{\text{BH}} = -\frac12{\mathcal Q}^t{\mathcal M}{\mathcal Q}\,.
\label{eq:vbh}
\end{equation}
As a consequence of \eqref{eq:Mprop}, it follows that the action of $\mathfrak{F}_z$ on $\mathcal Q$ is 
anti-involutive, $\mathfrak{F}_z^2(\mathcal Q) = -\mathcal{Q}$. Using again \eqref{eq:Mprop}, one
shows that
\begin{equation}
\mathfrak{F}_z (V_{\text{BH}}(\mathcal Q, z^i)) = -\frac12\hat{\mathcal{Q}}^t\mathcal{M}
\hat{\mathcal{Q}} = V_{\text{BH}}(\mathcal Q, z^i)\,, \label{eq:FV}
\end{equation}
i.e., the black hole potential is invariant under Freudenthal duality.
Moreover, the second equation of \eqref{eq:Mprop} yields
\begin{equation}
\partial_i\mathcal M = \mathcal{M}\Omega(\partial_i\mathcal{M})\Omega\mathcal M\,. \label{eq:Id}
\end{equation}
The direct application of this identity implies that under $\mathfrak{F}_z$, $\partial_i V_{\text{BH}}$ flips
sign\footnote{Since the operator $\mathfrak F_z$ does not commute with $\partial_i$, it is
important to specify that $\mathfrak F_z$ acts always after the action of $\partial_i$. Notice that
\eqref{eq:FVd} corrects eq.~(3.11) of \cite{Ferrara:2011gv}.},
\begin{equation}
\mathfrak F_z (\partial_i V_{\text{BH}}(\mathcal Q, z^i)) = -\frac12\hat{\mathcal Q}^t(\partial_i\mathcal M)
\hat{\mathcal Q} = -\partial_i V_{\text{BH}}(\mathcal Q, z^i)\,. \label{eq:FVd}
\end{equation}
Since the $z_{\text h}^i(\mathcal Q)$ are the critical points of $V_{\text{BH}}$, one has
\begin{equation}
0 = \partial_i V_{\text{BH}}|_{z_{\text h}^i(\mathcal Q)} = -\mathfrak F_z(\partial_i
V_{\text{BH}})|_{z_{\text h}^i(\mathcal Q)} = \frac12\hat{\mathcal Q}^t (\partial_i\mathcal M)
\hat{\mathcal Q}|_{z_{\text h}^i(\mathcal Q)} = \frac12\hat{\mathcal Q}_{\text h}^t\partial_i
\mathcal M (z_{\text h}^i(\mathcal Q))\hat{\mathcal Q}_{\text h}\,, \label{eq:criticalungauged}
\end{equation}
where we introduced Freudenthal duality $\mathfrak F$ at the horizon as
\begin{equation}
\mathfrak F(\mathcal Q) = \mathfrak F_z(\mathcal Q)|_{z_{\text h}^i(\mathcal Q)} = -\Omega
\mathcal M_{\text h}\mathcal Q = \hat{\mathcal Q}_{\text h}\,. \label{hor-limit}
\end{equation}
On the other hand, applying \eqref{eq:critun} to the charge configuration $\hat{\mathcal Q}_{\text h}$
leads to
\begin{equation}
0 = -\partial_i V_{\text{BH}}(\hat{\mathcal Q }_{\text h}, z^i)|_{z_{\text h}^i(\hat{\mathcal Q}_{\text h})}
= \frac12\hat{\mathcal Q }_{\text h}^t\partial_i\mathcal M (z_{\text h}^i(\hat{\mathcal Q}_{\text h}))
\hat{\mathcal Q}_{\text h}\,. \label{eq:dualcriticalungauged}
\end{equation}
Comparing \eqref{eq:criticalungauged} and \eqref{eq:dualcriticalungauged}, one can conclude that the
attractor configuration
\begin{equation}
z_{\text h}^i(\hat{\mathcal Q}_{\text h}) = z_{\text h}^i(\mathcal Q)\,,  \label{eq:stabsca}
\end{equation}
is a solution also for \eqref{eq:dualcriticalungauged} \cite{Ferrara:2011gv}. Eq.~\eqref{eq:stabsca} can be
interpreted as the stabilization of the near horizon configuration under Freudenthal duality, but an explicit
verification of this claim is possible only if all the charges are different from zero. In any case one can always
verify that $z^i_{\text h}$ is critical point for both $V_{\text{BH}}(\ma Q, z^i)$ and
$V_{\text{BH}}(\hat{\ma Q}_{\text h}, z^i)$.

This fact turns out to be crucial in order to extend \eqref{eq:actF} to a symmetry of the black hole entropy
$S_{\text{BH}}$. In fact, using \eqref{SBH}, \eqref{eq:FV} and \eqref{eq:stabsca}, one obtains
\begin{eqnarray}
\frac1{\pi}\mathfrak F (S_{\text{BH}}) &=& \mathfrak F\left(-\frac12\mathcal Q^t\mathcal M (z_{\text h}^i
(\mathcal Q))\mathcal Q\right) = -\frac12\hat{\mathcal Q}_{\text h}^t\mathcal M (z_{\text h}^i
(\hat{\mathcal Q}_{\text h}))\hat{\mathcal Q}_{\text h} \nonumber \\
&=& -\frac12\mathcal Q^t\mathcal M_{\text h}\mathcal Q = \frac{S_{\text{BH}}}\pi\,. \label{eq:Sinv}
\end{eqnarray}
Thus, the entropy pertaining to the charge configuration $\mathcal Q$ is the same as the one pertaining
to the Freudenthal dual configuration $\mathfrak F(\mathcal Q)$. Since $\mathfrak F(\mathcal Q)$
in \eqref{hor-limit} is homogeneous of degree one (but generally nonlinear) in $\mathcal Q$,
\eqref{eq:Sinv} results in the quite remarkable fact that the Bekenstein-Hawking entropy of a black hole in
ungauged supergravity is invariant under an intrinsically nonlinear map acting on charge configurations.
Note that no assumption has been made on the underlying special K\"ahler geometry, nor did we use
supersymmetry.

\subsection{\label{sec:FI}$\text{U}(1)$ FI-gauged $N=2$, $D=4$ supergravity}

In $\text{U}(1)$ FI-gauged $N=2$, $D=4$ supergravity, the parameters in terms of
which the scalars $z^i$ stabilize at the horizon, are doubled by the gauge
couplings $\mathcal G$. The entropy density and the horizon values of the scalars are now
determined by the effective potential \eqref{eq:effective}, which contains both $V_{\text{BH}}$
and the scalar potential $V$.

As a first step, we extend the action of the field-dependent Freudenthal duality $\mathfrak F_z$ by
acting on both $\mathcal Q$ and $\mathcal G$ according to
\begin{equation}
\mathfrak F_z (\mathcal Q) = \hat{\mathcal Q} = -\Omega\mathcal M\mathcal Q\,, \qquad
\mathfrak F_z (\mathcal G) = \hat{\mathcal G} = -\Omega\mathcal M\mathcal G\,, \label{eq:Fimages}
\end{equation}
while, by definition, $\mathfrak F_z$ leaves the symplectic section $\mathcal V$ (and its covariant derivatives) invariant. Now use \eqref{eq:Mprop}, \eqref{eq:Id}, and the fact that the scalar potential
can be written as \cite{Dall'Agata:2010gj,Klemm:2016wng}
\begin{equation}
V = g^{i\bar\jmath} D_i\mathcal L\bar D_{\bar\jmath}\bar{\mathcal L} - 3\vert\mathcal L\vert^2
= -\frac12\mathcal G^t\mathcal M\mathcal G - 4\vert\mathcal L\vert^2\,, \label{Vg}
\end{equation}
where
\begin{equation}
\mathcal L\equiv\mathcal G^t\Omega\mathcal V = \langle\mathcal G,\mathcal V\rangle\,, \label{L-call}
\end{equation}
to obtain
\begin{equation}
\begin{split}
&\mathfrak F_z (V(\mathcal G, z^i)) = -\frac12\hat{\mathcal G}^t\mathcal M\hat{\mathcal G}
- 4\hat{\mathcal L}\hat{\bar{\mathcal L}} = V(\mathcal G, z^i)\,, \\
&\mathfrak F_z (\partial_i V(\mathcal G, z^i)) = -\frac12\hat{\mathcal G}^t (\partial_i\mathcal M) 
\hat{\mathcal G} - 4(D_i\hat{\mathcal L})\hat{\bar{\mathcal L}} = -\partial_i V(\mathcal G, z^i)\,.
\end{split} \label{July}
\end{equation}
Since $V_{\text{eff}}$ and $\partial_i V_{\text{eff}}$ (cf.~(2.26) of \cite{Bellucci:2008cb}) can be written
as functions of $V_{\text{BH}}$, $V$, $\partial_i V_{\text{BH}}$ and $\partial_i V$, \eqref{July},
together with \eqref{eq:FV} and \eqref{eq:FVd} implies
\begin{equation}
\mathfrak F_z (V_{\text{eff}}(\mathcal Q, \mathcal G, z^i)) = V_{\text{eff}}(\mathcal Q, \mathcal G, z^i)\,,
\qquad\mathfrak F_z (\partial_i V_{\text{eff}}(\mathcal Q, \mathcal G, z^i)) = -\partial_i V_{\text{eff}}
(\mathcal Q, \mathcal G, z^i)\,. \label{eq:FVgge}
\end{equation}
Using the second relation of \eqref{eq:FVgge}, one has then
\begin{eqnarray}
0& =& -\partial_i V_{\text{eff}}|_{z_{\text h}^i(\mathcal Q, \mathcal G)} = \mathfrak F_z (\partial_i
V_{\text{eff}})|_{z_{\text h}^i(\mathcal Q, \mathcal G)} \nonumber \\
&=&\partial_i V_{\text{eff}}(\hat{\mathcal Q}, \hat{\mathcal G}, z^i)|_{z_{\text h}^i (\mathcal Q,
\mathcal G)} = \partial_i V_{\text{eff}}(\hat{\mathcal Q}_{\text h}, \hat{\mathcal G}_{\text h}, z_{\text h}^i
(\mathcal Q, \mathcal G))\,. \label{eq:Feffective}
\end{eqnarray}
Let us define Freudenthal duality at the horizon by
\begin{eqnarray}
\mathfrak F (\mathcal Q) &=& \mathfrak F_z (\mathcal Q)|_{z_{\text h}^i(\mathcal Q, \mathcal G)}
= -\Omega\mathcal M_{\text h}\mathcal Q =\hat{\mathcal Q}_{\text h}\,, \nonumber \\
\mathfrak F (\mathcal G) &=& \mathfrak F_z (\mathcal G)|_{z_{\text h}^i(\mathcal Q, \mathcal G)}
= -\Omega\mathcal M_{\text h}\mathcal G =\hat{\mathcal G}_{\text h}\,. \label{eq:hF}
\end{eqnarray}
From the comparison of \eqref{eq:Feffective} with the definition
\begin{equation}
0 = \partial_i V_{\text{eff}}(\hat{\mathcal Q}_{\text h}, \hat{\mathcal G}_{\text h}, z^i)|_{z_{\text h}^i
(\hat{\mathcal Q}_{\text h}, \hat{\mathcal G}_{\text h})} = \partial_i V_{\text{eff}}(\hat{\mathcal Q}_{\text h},
\hat{\mathcal G}_{\text h}, z_{\text h}^i (\hat{\mathcal Q}_{\text h}, \hat{\mathcal G}_{\text h}))\,,
\label{eq:dualderveff}
\end{equation}
it follows that
\begin{equation}
z_{\text h}^i (\hat{\mathcal Q}_{\text h}, \hat{\mathcal G}_{\text h}) = z_{\text h}^i (\mathcal Q,
\mathcal G) \label{eq:stabscag}
\end{equation}
is a solution also for \eqref{eq:dualderveff}, thus it is a critical point for both $V_{\text{eff}}$ and
$\maf F (V_{\text{eff}})$.

Eqns.~\eqref{entr-density}, \eqref{eq:FVgge} and \eqref{eq:stabscag} imply that $s_{\text{BH}}$ is
invariant under Freudenthal duality,
\begin{eqnarray}
4\mathfrak F (s_{\text{BH}}) &=& V_{\text{eff}}(\hat{\mathcal Q}_{\text h}, \hat{\mathcal G}_{\text h},
z_{\text h}^i (\hat{\mathcal Q}_{\text h}, \hat{\mathcal G}_{\text h})) =
V_{\text{eff}}(\hat{\mathcal Q}_{\text h}, \hat{\mathcal G}_{\text h}, z_{\text h}^i (\mathcal Q,\mathcal G))
\nonumber \\
&=& V_{\text{eff}}(\mathcal Q, \mathcal G, z_{\text h}^i (\mathcal Q, \mathcal G)) =
4 s_{\text{BH}}\,. \label{eq:Sinvg}
\end{eqnarray}
It is immediate to see that in the limit $\mathcal G\to 0$, one recovers the results of the ungauged case.
Notice that the origin of Freudenthal duality is firmly rooted into the properties \eqref{eq:Mprop}. The
action of $\mathfrak F$ yields a new attractor-supporting configuration $(\hat{\mathcal Q}_{\text h},
\hat{\mathcal G}_{\text h})$ that, in general, belongs to a physically different theory, specified by a
different choice of gauge couplings.

It is worthwhile to note that no assumption has been made on the special K\"ahler geometry of the
scalars in the vector multiplets. The invariance \eqref{eq:Sinvg} holds thus also in models with
non-homogeneous special K\"ahler manifolds, like e.g.~the quantum stu model recently treated
in \cite{Klemm:2015xda}.

As an illustrative example, let us check the action of Freudenthal duality for the simple model with
prepotential $F=-i X^0 X^1$ and purely electric FI gauging, cf.~\cite{Cacciatori:2009iz} for
details\footnote{As discussed in sec.~10 of \cite{Ferrara:2012qp}, the Freudenthal duality of $N=2$,
$D=4$ supergravity minimally coupled to a certain number of vector multiplets in the ungauged case is 
nothing but a particular anti-involutive symplectic transformation of the U-duality.}.
To keep things simple, we assume that the electric charges vanish. One has thus
\begin{equation}
\ma Q = \left(\begin{array}{c}
p^0 \\
p^1\\
0\\
0
\end{array}\right)\,, \qquad \ma G = \left(\begin{array}{c}
0 \\
0 \\
g_0 \\
g_1
\end{array}\right)\,.
\end{equation}
This model has just one complex scalar $z=x+iy$, and the matrix $\mathcal M$ is given by
\begin{equation}
\mathcal M = \left(\begin{array}{cccc} -\frac{x^2 + y^2}x & 0 & \frac yx & 0 \\
0 & -\frac1x & 0 & -\frac yx \\
\frac yx & 0 & -\frac1x & 0 \\
0 & -\frac yx & 0 & -\frac{x^2 + y^2}x
\end{array}\right)\,.
\end{equation}
The black hole and scalar potential read respectively
\begin{eqnarray}
V_{\text{BH}} &=& -\frac12\mathcal Q^t\mathcal M\mathcal Q = \frac{x^2 + y^2}{2x}(p^0)^2 +
\frac{(p^1)^2}{2x}\,, \nonumber \\
V &=& -\frac1{2x} (g_0^2 + 4g_0 g_1 x + g_1^2 (x^2 + y^2))\,. \label{VVBHX0X1}
\end{eqnarray}
Plugging this into the effective potential \eqref{eq:effective}, one shows that the latter is extremized for
\begin{equation}
x = x_{\text h} = \frac{u g_0}{g_1}\,, \qquad y = y_{\text h} = 0\,, \label{eq:X0X1}
\end{equation}
where $u$ is a solution of the quartic equation
\begin{equation}
\left[(1 - \nu^2) u + 2(u^2 - \nu^2)\right]^2 = k(1 - u^2)(\nu^2 - u^2)\,, \label{eq:u-quartic}
\end{equation}
with
\begin{equation}
\nu\equiv\frac{g_1 p^1}{g_0 p^0}\,, \qquad k\equiv\frac{\kappa^2}{(g_0 p^0)^2}\,.
\end{equation}
Note that positivity of the kinetic terms in the action requires $x>0$. Depending on the sign of
$g_0/g_1$, this means that either only negative or only positive roots of \eqref{eq:u-quartic} are
allowed, and such roots may not exist for all values of $\nu$ and $k$. Notice also that in the special
case where
\begin{equation}
(2g_0 p^0)^2 = (2g_1 p^1)^2 = \kappa^2\,, \label{eq:cond-BPS-sign}
\end{equation}
the effective potential \eqref{eq:effective} becomes completely flat,
\begin{equation}
V_{\text{eff}} = -\frac{\kappa}{2 g_0 g_1}\,,
\end{equation}
and the scalar $z$ is thus not stabilized at the horizon, a fact first noted in \cite{Cacciatori:2009iz}.
(Nonetheless, the entropy is still independent of the arbitrary value $z_{\text h}$, in agreement with
the attractor mechanism). \eqref{eq:cond-BPS-sign} corresponds to the BPS conditions
found in \cite{Cacciatori:2009iz}, or to a sign-flipped modification of them\footnote{In the BPS case,
$g_0p^0$ and $g_1p^1$ must have the same sign.}. It would be interesting to see whether the
appearance of flat directions is a generic feature of the BPS case, or just a consequence of the
simplicity of the model under consideration. A large class of supersymmetric black holes in gauged 
supergravity satisfies a Dirac-type quantization condition \cite{Cacciatori:2009iz} (that corresponds to a 
twisting of the dual superconformal field theory \cite{Maldacena:2000mw}), i.e., one has a relation between
$\ma Q$ and $\ma G$, that enter into $V_{\text{BH}}$ and $V$ respectively. This indicates that
flat directions of $V_{\text{eff}}$ might be generic in the supersymmetric case.

Using \eqref{eq:u-quartic}, one can derive the near-horizon value of $V_{\text{eff}}$, and thus the entropy
density \eqref{entr-density},
\begin{equation}
s_{\text{BH}} = \frac{V_{\text{eff}}(\mathcal Q, \mathcal G, z^i)|_{z_{\text h}^i(\mathcal Q, \mathcal G)}}4
= \frac{g_0 {p^0}^2 [(1 - \nu^2) u + 2(u^2 - \nu^2)]}{4\kappa g_1 (1 - u^2)}\,.
\label{eq:entropyX0X1}
\end{equation}
We now determine the action of Freudenthal duality on the charges and the FI parameters.
The definitions \eqref{eq:hF} yield
\begin{equation}
\maf F(\ma Q)\equiv\left(\begin{array}{c} 0 \\ 0 \\ \hat{q}_0 \\ \hat{q}_1\end{array}\right) =
\left(\begin{array}{c} 0 \\ 0 \\ p^0 x_{\text h} \\ p^1/x_{\text h}\end{array}\right)\,, \qquad
\maf F(\ma G)\equiv\left(\begin{array}{c} \hat{g}^0 \\ \hat{g}^1 \\ 0 \\ 0\end{array}\right) =
\left(\begin{array}{c} -g_0/x_{\text h} \\ -g_1 x_{\text h} \\ 0 \\ 0\end{array}\right)\,.
\label{eq:dualchargesX0X1}
\end{equation}
The dual configuration is thus electrically charged and has purely magnetic gaugings.
For the transformed potentials one gets
\begin{eqnarray}
\maf F (V_{\text{BH}}) &=& -\frac12\hat{\mathcal Q}^t_{\text h}\mathcal M\hat{\mathcal Q}_{\text h} =
\frac{x^2 + y^2}{2x}\hat{q}_1^2 + \frac{\hat{q}_0^2}{2x}\,, \\
\maf F (V) &=& -\frac12\hat{\mathcal G}^t_{\text h}\mathcal M\hat{\mathcal G}_{\text h} - 4|\langle
\hat{\mathcal G}_{\text h}, \mathcal V\rangle|^2
= -\frac1{2x}\left((\hat{g}^1)^2 + 4\hat{g}^0\hat{g}^1 x + (\hat{g}^0)^2 (x^2 + y^2)\right)\,. \nonumber
\end{eqnarray}
These are identical to \eqref{VVBHX0X1}, except for the replacements
\begin{displaymath}
(p^0)^2 \to \hat{q}_1^2\,, \qquad (p^1)^2 \to \hat{q}_0^2\,, \qquad g_0^2 \to (\hat{g}^1)^2\,,
\qquad g_1^2 \to (\hat{g}^0)^2\,, \qquad g_0 g_1 \to \hat{g}^0\hat{g}^1\,.
\end{displaymath}
The critical points of $\maf F(V_{\text{eff}})$ are thus $\hat{x}_{\text h}=\hat{g}^1\hat{u}/\hat{g}^0$
and $\hat{y}_{\text h}=0$, where $\hat u$ satisfies
\begin{equation}
\left[(1 - \hat{\nu}^2)\hat u + 2(\hat{u}^2 - \hat{\nu}^2)\right]^2 = \hat k(1 - \hat{u}^2)(\hat{\nu}^2
- \hat{u}^2)\,, \label{eq:hatu-quartic}
\end{equation}
with
\begin{equation}
\hat\nu\equiv\frac{\hat{g}^0\hat{q}_0}{\hat{g}^1\hat{q}_1}\,, \qquad\hat k\equiv
\frac{\kappa^2}{(\hat{g}^1\hat{q}_1)^2}\,.
\end{equation}
Now, using \eqref{eq:dualchargesX0X1}, one easily shows that
\begin{displaymath}
\hat{\nu}^2 = \frac1{\nu^2}\,, \qquad \hat k = \frac k{\nu^2}\,.
\end{displaymath}
Plugging this into \eqref{eq:hatu-quartic} and multiplying with $\nu^4/\hat{u}^4$ yields
\begin{equation}
\left[(1 - \nu^2)\hat{u}^{-1} + 2(\hat{u}^{-2} - \nu^2)\right]^2 = k(1 - \hat{u}^{-2})(\nu^2 - \hat{u}^{-2})\,.
\end{equation}
Comparing with \eqref{eq:u-quartic}, we see that $u$ and $\hat{u}^{-1}$ satisfy the same equation,
and have thus the same set of solutions. Hence, up to permutations of possible multiple roots, one gets
$u=\hat{u}^{-1}$, which, by means of \eqref{eq:dualchargesX0X1}, leads to $\hat x_{\text h}=x_{\text h}$,
and therefore $V_{\text{eff}}$ and $\maf F(V_{\text{eff}})$ share the same critical points.

The transformed entropy density is given by
\begin{equation}
\maf F(s_{\text{BH}}) = \frac{V_{\text{eff}}(\maf F(\mathcal Q), \maf F(\mathcal G),z^i)|_{\hat{
z}_{\text h}^i(\maf F(\mathcal Q), \maf F(\mathcal G))}}4 = \frac{\hat{g}^1\hat{q}_1^2 [(1 - \hat{\nu}^2)
\hat u + 2(\hat{u}^2 - \hat{\nu}^2)]}{4\kappa\hat{g}^0 (1 - \hat{u}^2)}\,.
\end{equation}
Using again \eqref{eq:dualchargesX0X1}, it is easy to see that this coincides with \eqref{eq:entropyX0X1},
so that the entropy is indeed invariant under Freudenthal duality.

\subsection{Coupling to hypermultiplets}
\label{hypers}

In this section we generalize our analysis to include also hypermultiplets, and consider the case
where abelian isometries of the quaternionic hyperscalar target manifold are gauged.
The dynamics of the attractor mechanism is now governed by the potentials $V_{\text{BH}}(\ma Q, z^i)$
and $V(\ma P^x(q^u),{\mathcal K}^u,z^i)$, where $\ma P^x=(\ma P^{x\Lambda},\ma P^x_\Lambda)$
denote the triholomorphic moment maps, and $\mathcal K^u=(k^{\Lambda u},k^u_\Lambda)$ are
the Killing vectors that define the gauging. Note the presence of magnetic moment maps
$\ma P^{x\Lambda}$ and magnetic Killing vectors $k^{\Lambda u}$.
In what follows, we introduce the collective index $A=(i,u)$ and represent the scalars as
\begin{equation}
\phi^A = (z^i, q^u)\,.
\end{equation}
As was shown in \cite{Klemm:2016wng}, the scalar potential can be written in the symplectically
covariant form
\begin{equation}
V = \mathbb{G}^{AB}\mathbb{D}_A\mathcal L\,\mathbb{D}_B\bar{\mathcal L} - 3|\mathcal L|^2\,,
\end{equation}
where
\begin{displaymath}
\mathbb{G}^{AB} = \left(\begin{array}{cc} g^{i\bar{\jmath}} & 0 \\ 0 & h^{uv}\end{array}\right)\,, \qquad
\mathbb{D}_A = \left(\begin{array}{c} D_i \\ \masf{D}_u\end{array}\right)\,, \qquad \mathcal L =
\langle\ma Q^x\ma P^x, \mathcal V\rangle\,, \qquad \ma Q^x = \langle\ma P^x, \ma Q\rangle\,,
\end{displaymath}
provided the `quantization condition' $\ma Q^x\ma Q^x=1$ holds\footnote{This represents a rather mild
assumption, cf.~footnote 8 of \cite{Klemm:2016wng}.}.

The field-dependent Freudenthal duality is again defined by \eqref{eq:actF}, supplemented with
\begin{equation}
\mathfrak F_z (\mathcal P^x)\equiv\hat{\mathcal P}^x = -\Omega\mathcal M\mathcal P^x\,,
\qquad\mathfrak F_z (\mathcal K^u )\equiv\hat{\mathcal K}^u = -\Omega\mathcal M\mathcal K^u\,.
\end{equation}
One easily shows that $\mathfrak F_z(\ma Q^x)=\ma Q^x$ and, with slightly more effort, that
\begin{equation}
\begin{split}
&\mathfrak F_{z} (V_{\text{eff}}(\ma Q,\ma P^x (q^u), \mathcal K^u (q^u), z^i)) = V_{\text{eff}}
(\ma Q,\ma P^x (q^u), \mathcal K^u (q^u), z^i)\,, \\
&\mathfrak F_{z} (\partial_A V_{\text{eff}}(\ma Q, \ma P^x (q^u), \mathcal K^u (q^u), z^i)) =
-\partial_A V_{\text{eff}}(\ma Q, \ma P^x (q^u), \mathcal K^u (q^u), z^i)\,.
\label{eq:invarianceVeffhyper}
\end{split}
\end{equation}
Thus, in analogy to the $\text{U}(1)$ FI case, one has to consider the criticality conditions \eqref{eq: crithyp}
and apply the second relation of \eqref{eq:invarianceVeffhyper},
\begin{equation}
\begin{split}
0 & = -\partial_A V_{\text{eff}}(\ma Q, \mathcal P^x, \mathcal K^u, z^i)|_{\phi_{\text h}^A} =
\maf F_z(\partial_A V_{\text{eff}}(\ma Q, \mathcal P^x, \mathcal K^u, z^i))|_{\phi_{\text h}^A} = \\
& = \partial_A V_{\text{eff}}(\hat{\ma Q}, \hat{\mathcal P}^x, \hat{\mathcal K}^u, z^i)|_{\phi_{\text h}^A}
= \partial_A V_{\text{eff}}(\hat{\ma Q}_{\text h}, \hat{\mathcal P}^x_{\text h} (q^u_{\text h}),
\hat{\mathcal K}^u (q^u_{\text h}), z^i_{\text h})\,, \label{eq:criticalhyper}
\end{split}
\end{equation}
where
\begin{equation}
\hat{\mathcal P}^x_{\text h} (q^u) = -\Omega\mathcal M_{\text h}\mathcal P^x( q^u)
\end{equation}
is the dual expression for the moment maps that depends on the scalar fields, the charges and the 
parameters contained in the quaternionic Killing vectors.
Defining $\hat{\ma Q}_{\text h}$ as in \eqref{eq:hF}, the criticality condition of the attractor points
$\hat{\phi}_{\text h}^A$ for the dual configuration of $(\ma Q, \ma P^x(q^u))$, namely for
$(\hat{\ma Q}_{\text h},\hat{\mathcal P}_{\text h}^x(q^u))$, reads
\begin{equation}
0 = \partial_A V_{\text{eff}}(\hat{\ma Q}_{\text h}, \hat{\mathcal P}_{\text h}^x, \hat{\mathcal K}^u,
z^i)|_{\hat{\phi}_{\text h}^A} = \partial_A V_{\text{eff}}(\hat{\ma Q}_{\text h}, \hat{\mathcal P}_{\text
h}^x (\hat q^u_{\text h}), \hat{\mathcal K}^u (\hat q^u_{\text h}), \hat z^i_{\text h})\,.
\label{eq:dualcriticalhyper}
\end{equation}
Thus a comparison between \eqref{eq:criticalhyper} and \eqref{eq:dualcriticalhyper} shows that the
configuration
\begin{equation}
\phi_{\text h}^A = \hat{\phi}_{\text h}^A
\end{equation}
is a solution for both criticality conditions. It follows that
\begin{equation}
\begin{split}
4\maf F (s_{\text{BH}}) = & V_{\text{eff}}(\hat{\ma Q}_{\text h}, \hat{\mathcal P}^x_{\text h}
(\hat{q}^u_{\text h}), \hat{z}_{\text h}^i) = V_{\text{eff}}(\hat{\ma Q}_{\text h},\hat{\mathcal P}^x_{\text h}
(q^u_{\text h}), z^i_{\text h}) \\
& = V_{\text{eff}}(\ma Q, \mathcal P^x_{\text h} (q^u_{\text h}), z^i_{\text h}) = 4 s_{\text{BH}}\,,
\label{eq:Sinvhyper}
\end{split}
\end{equation}
namely the entropy density of the two configurations related by the Freudenthal operator is the same.

\section{\label{conclusion}Final remarks}

In this paper, Freudenthal duality, a nonlinear anti-involutive map
defined on the electromagnetic charge representation space of Einstein-Maxwell systems
coupled to non-linear sigma models, was extended to the case where
abelian isometries of $N=2$, $D=4$ supergravity coupled to vector- and hypermultiplets are gauged.

Without any assumption on the geometry of the scalar manifolds, the
Bekenstein-Hawking entropy was shown to be invariant under such a
nonlinear map, which generally commutes with local supersymmetry (if any). As far as we know,
this is the first example of a nonlinear symmetry of the black hole entropy itself, whose general
invariance is usually given by the electromagnetic symplectic duality
transformations, which act linearly on the charges and on the gauging
parameters (within a symplectically covariant formalism).

Many further developments are possible, along the lines of the present
investigation. We list and comment on some of them below.

As we pointed out, Freudenthal duality does not need supersymmetry, even if it was
originally introduced in \cite{Borsten:2009zy} in the context of $D=4$ supergravity
theories with symmetric scalar manifolds. Along this venue of research, it
would be interesting to extend the results presented above to abelian
gaugings in theories with extended ($N>2$) supersymmetry,
and also to certain classes of $N=1$ models, whose symplectic structure is compatible with
electromagnetic duality, thus allowing for an attractor behaviour of the near-horizon dynamics of the
scalar flow (cfr.~e.g.~\cite{Andrianopoli:2007rm}).

Since its introduction, the stringy origin of Freudenthal duality has always
remained a mistery, with its nonlinearity hinting to a nonperturbative
nature. Also in view of the extension to the presence of gaugings - which generally characterize
the supergravity theories obtained as low-energy limit of string and M-theory compactifications -, it
would be interesting to deal with the challenging task of a
realization of the Freudenthal anti-involutive map in higher-dimensional
string/M theory.

Finally, one can try to analyze the role and meaning of the intrinsically non-linear map provided by
Freudenthal duality in string/M-theory flux compactifications, by
using the AdS/CFT correspondence, especially in relation to recent
results in which the large $N$ partition function of ABJM theory on spaces of the type
$\Sigma\times\text{S}^1$ was shown to reproduce the Bekenstein-Hawking entropy of
static AdS$_4$ BPS black holes \cite{Benini:2015eyy}.
Moreover, the role of the attractor mechanism for static black hole solutions in
gauged supergravity coupled to hypermultiplets deserves further investigations. In particular,
for both the models proposed in \cite{Martelli:2009ga} and \cite{Jafferis:2009th} to be dual to
$\text{AdS}_4\times V^{5,2}/\mathbb Z_k$, the field theory computations \cite{Hosseini:2016ume} show
the same value of the topological free energy (up to a linear affine transformation of chemical potentials
and charges). This fact may be related to Freudenthal duality and might point to some hidden link between
the topological free energy and black hole entropy.

\section*{Acknowledgements}

A.~M.~would like to thank the Physics Dept.~of the University of Milan for kind
hospitality and stimulating environment.

\end{document}